\documentclass[letterpaper]{article} % DO NOT CHANGE THIS
\usepackage{aaai23}  % DO NOT CHANGE THIS
\usepackage{times}  % DO NOT CHANGE THIS
\usepackage{helvet}  % DO NOT CHANGE THIS
\usepackage{courier}  % DO NOT CHANGE THIS
\usepackage[hyphens]{url}  % DO NOT CHANGE THIS
\usepackage{graphicx} % DO NOT CHANGE THIS
\usepackage{natbib}  % DO NOT CHANGE THIS AND DO NOT ADD ANY OPTIONS TO IT
\usepackage{caption} % DO NOT CHANGE THIS AND DO NOT ADD ANY OPTIONS TO IT
\usepackage{algorithm}
\usepackage{algorithmic}
\usepackage{amsmath}
\usepackage{booktabs}
\usepackage{amsfonts}

\usepackage{newfloat}
\usepackage{listings}
\DeclareCaptionStyle{ruled}{labelfont=normalfont,labelsep=colon,strut=off} % DO NOT CHANGE THIS
\floatstyle{ruled}
\newfloat{listing}{tb}{lst}{}
\floatname{listing}{Listing}
\title{Can Adversarial Networks Make Uninformative Colonoscopy Video Frames Clinically Informative? (Student Abstract)}
\author{
    Vanshali Sharma,
    M.K. Bhuyan,
    Pradip K. Das
}
\affiliations{
    Indian Institute of Technology Guwahati, Assam, India-781039\\
     \{vanshalisharma, mkb, pkdas\}@iitg.ac.in
}

\begin{document}

\maketitle

\begin{abstract}
Various artifacts, such as ghost colors, interlacing, and motion blur, hinder diagnosing colorectal cancer (CRC) from videos acquired during colonoscopy. The frames containing these artifacts are called uninformative frames and are present in large proportions in colonoscopy videos. To alleviate the impact of artifacts,  we propose an adversarial network based framework to convert uninformative frames to clinically relevant frames. We examine the effectiveness of the proposed approach by evaluating the translated frames for polyp detection using YOLOv5. Preliminary results present improved detection performance along with elegant qualitative outcomes. We also examine the failure cases to determine the directions for future work.
\end{abstract}

\section{Introduction}

Colonoscopy is a minimally invasive procedure widely adopted for polyp detection to diagnose colorectal cancer (CRC). In a colonoscopy, diagnostic accuracy relies on the correct analysis of the acquired recordings. However, the traditional assessment approaches by physicians suffer from inter-observer variations and demand extensive manual efforts. In recent years, accessibility to several colonoscopy datasets has paved the way for many machine learning based research works for automated CRC detection. However, the well-trained models proposed in the existing works still report limited diagnostic success. This limited success of automated methods is attributed to low-quality frames in the video samples, which contain artifacts, namely, ghost colors, low-illumination, interlacing due to camera motion, and fecal depositions due to inadequate patient preparation. 

\par To overcome the low-quality frames, some related fields of laparoscopy and endoscopy followed keyframe selection \cite{ma2020keyframe} or performed super-resolution \cite{almalioglu2020endol2h}, but no work in the colonoscopy domain explored the idea of extracting obscured clinical details from such low-quality uninformative video frames. Therefore, our work investigates whether GANs can convert uninformative frames to informative frames. In this direction, we propose a GAN-based image-to-image translation approach to generate informative frames from the degraded frames of the colonoscopy videos. We highlight the cases where GANs fail and where it helps, which gives us directions for future work.
The main contributions are summarized below:

\begin{enumerate}
    \item To the best of our knowledge, this is the first framework to address the issue of uninformative colonoscopy frames using adversarial networks.
    \item We investigate the impact of translating uninformative frames on polyp detection performance and discuss future directions in this context.
\end{enumerate}

\begin{figure}
    \centering
    \includegraphics[height=130pt, width=\columnwidth]{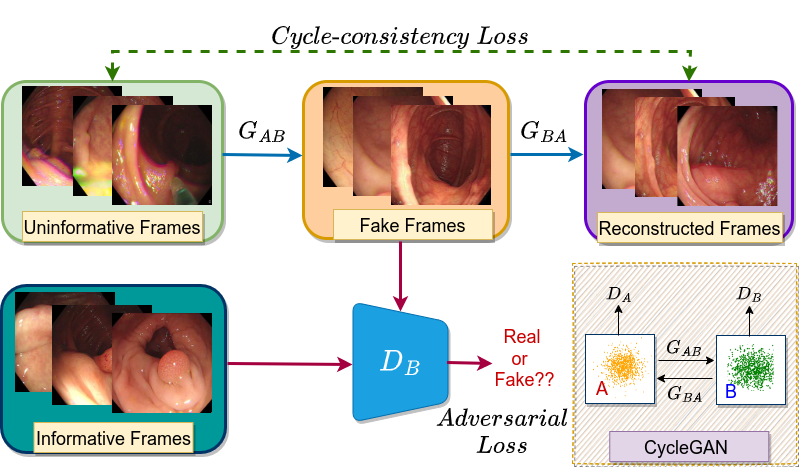}
    \caption{The proposed framework contains two generators $G_{AB}$ and $G_{BA}$ and two discriminators $D_A$ and $D_B$.}
    \label{fig:gan}
\end{figure}

\section{Methodology}
The overview of the proposed framework is shown in Fig. \ref{fig:gan}. Given the uninformative colonoscopy frames $\{a_i\}_{i=1}^N$ from domain A, the aim is to learn a mapping function $G_{AB}: A \rightarrow B$ to generate frames such that the data distribution of obtained frames is indistinguishable from that of informative colonoscopy frames $\{b_j\}_{j=1}^M$ of domain B. Due to the unavailability of paired data, our work is inspired by the unpaired translation approach of CycleGAN \cite{zhu2017unpaired}. Hence, another mapping function $G_{BA}: B \rightarrow A$ is also introduced. Our implementation involves ResNet-based generators and PatchGAN discriminators $D_A$ and $D_B$. The CycleGAN objective integrates adversarial loss and cycle-consistency loss. The adversarial loss can be expressed as:

\begin{equation}
\label{adv}
  \begin{split}
    L_{adv}(G_{AB}, D_B) &= \mathbb{E}_{b\sim p_{data}(b)}[(D_B (b)-1)^2] \\ & + \mathbb{E}_{a\sim p_{data}(a)}[(D_B(G_{AB}(a)))^2]
     \end{split}
\end{equation}
$G_{AB}$ aims to translate uninformative frames such that they appear similar to the informative frames, while $D_B$ tries to distinguish the translated frames from the high-quality, informative frames of domain B. In other words, $D_B$ is trained to minimize $L_{adv}(G_{AB}, D_B)$ and $G_{AB}$ is trained to minimize $\mathbb{E}_{a\sim p_{data}(a)}[(D_B(G_{AB}(a))-1)^2]$.

To ensure cycle-consistency and to reduce randomness in mapping, a cycle-consistency loss is used, which is given by:  
\begin{equation} 
   \begin{split}
    L_{cyc}(G_{AB}, G_{BA}) & = \mathbb{E}_{a\sim p_{data}(a)}[\lVert G_{BA}(G_{AB}(a))-a\rVert_1] \\ & + \mathbb{E}_{b\sim p_{data}(b)}[\lVert G_{AB}(G_{BA}(b))-b\rVert_1]
  \end{split}
\end{equation}
An identity mapping loss is also added to help preserve color in translated images. 
With this model, we intend to determine the clinically relevant details obscured by the artifacts. Furthermore, we carried out the following investigations:
\begin{enumerate}
    \item Polyp detection is performed using YOLOv5 \cite{yolov5} to determine the impact of GAN-translated frames.
    \item Qualitative analysis is done to identify the artifacts successfully handled by the CycleGAN and analyze the ones that still persist in the translated frames.
\end{enumerate}

\section{Experiments}
\begin{figure}
 \centering
 \large
  \resizebox{\columnwidth}{!}{
    \begin{tabular}{cccc}
          \includegraphics[width=0.28\textwidth,height=130pt]{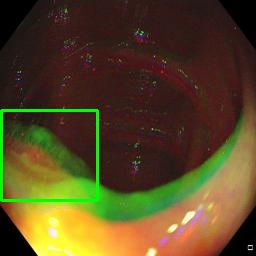} \hfill
          \includegraphics[width=0.28\textwidth,height=130pt]{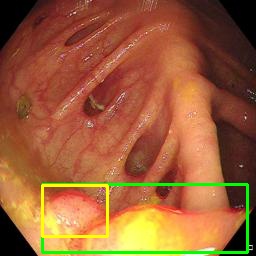} \hfill
          \includegraphics[width=0.28\textwidth,height=130pt]{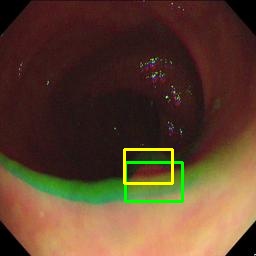} \hfill
          \includegraphics[width=0.28\textwidth,height=130pt]{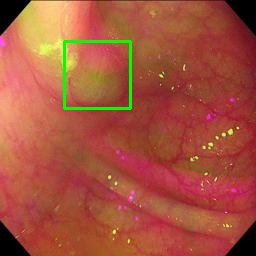}
           \\
            (a) 
          \\
          
          \includegraphics[width=0.28\textwidth,height=160pt]{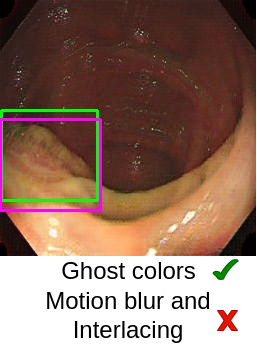}  \hfill
          \includegraphics[width=0.28\textwidth,height=160pt]{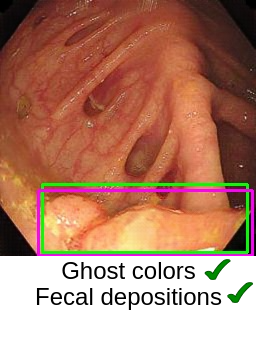}  \hfill
          \includegraphics[width=0.28\textwidth,height=160pt]{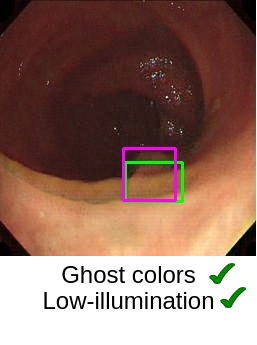} \hfill
          \includegraphics[width=0.28\textwidth,height=160pt]{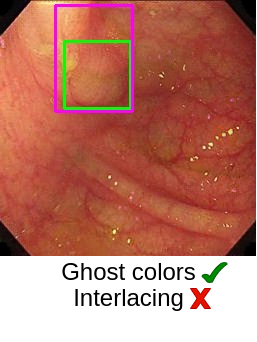} \\
           (b)

    \end{tabular}
    }
     \caption{ Detection performance using: (a) Raw frames and (b) Translated frames. Green bounding boxes denote the ground truth. Ticks and cross marks represent the successful and unsuccessful artifacts translations, respectively.
     }
      \label{fig:sample_res}
      
\end{figure}

To assess the effectiveness of the adversarial approach in mitigating the impact of artifacts, we conducted experiments using a publicly available SUN database \cite{misawa2021development} consisting of $1,09,554$ non-polyp and $49,136$ polyp frames. In addition to the localization information, the polyp frames are manually annotated by experts as informative or uninformative. We used only the polyp frames with a patient-wise split. 
The translation is done on a Titan Xp GPU at 14 frames per second.
%ratio of approximately 80, 10, 10 for training, validation, and test set, respectively. 
We report the results based on visual perception and consider feature space representation by evaluating the polyp detection outcomes using YOLOv5. We conducted training and testing in two scenarios using: a) Raw frames comprising both high and low-quality frames and b) Translated frames along with high-quality frames. The results in Table \ref{tab:localization} show that the translated frames complement the detection ability of YOLOv5 in terms of precision, recall, F1-score, and mAP@0.5. The detector correctly identified more polyps with lower deviations, presenting a more robust model. However, this is achieved with slightly less precise bounding boxes, as indicated by a minor decrease in mAP@0.5:0.95. Fecal depositions, ghost colors, and low-illumination are significantly reduced using CycleGAN, as shown in Fig. \ref{fig:sample_res}. However, motion blur and interlacing are not handled adequately in the process. This could be overcome by adopting blur removal approaches.

\begin{table}

    \centering
    \large
    \resizebox{0.85\columnwidth}{!}{
    \begin{tabular}{ccc}
    
      \toprule % <-- Toprule here
          & \multicolumn{2}{c}{\textbf{SUN Database}}  \\
      \cline{2-3} 
      \rule{0pt}{3ex}   
        \textbf{Metrics} & \textbf{Raw Frames} & \textbf{Translated Frames} \\
      \midrule
      
      \textbf{Precision (\%)} &
      92.03$\pm$0.60 & \textbf{93$\pm$0.87} \\
      
      \textbf{Recall (\%)} &
      88.9$\pm$3.12 &
      \textbf{90.2$\pm$1.3} \\

      \textbf{F1-score (\%)} &
      90.4$\pm$1.51 &
      \textbf{91.57$\pm$0.38} \\
      
      \textbf{mAP@0.5 (\%)} &
      95.37$\pm$0.95 &
      \textbf{95.6$\pm$0.21} \\
      
      \textbf{mAP@0.5:0.95 (\%)} &
      \textbf{57.53$\pm$0.32} &
      57.07$\pm$0.31 \\
     
      \bottomrule % <-- Bottomrule here
    \end{tabular}
    }

     \caption{Comparative analysis of polyp detection results}
     \label{tab:localization}
\end{table}

\section{Conclusion and Future Work}
In this work, we propose a GAN-based framework to translate uninformative colonoscopy frames into clinically significant frames. We showed that the translated frames improve polyp detection F1-score and mAP@0.5, with negligible reduction in mAP@0.5:0.95. We analyzed the types of artifacts where the CycleGAN performed well and identified the scope of improvements. Since the artifacts in colonoscopy video frames alter the various aspects of images, such as structure, texture, and color, this work lays the foundation for a more interesting future work of developing a standalone model to address all the artifacts in one go.

\section{Acknowledgments}

Vanshali Sharma would like to thank the Department of Science and Technology, Government of India, for providing the INSPIRE fellowship (IF190362).

\fontsize{9.0pt}{10.0pt} \selectfont

%\bibentry{c:23}

\bibliography{aaai23}

\end{document}